\documentclass[a4paper,11pt]{article}

\pdfoutput=1 % if your are submitting a pdflatex (i.e. if you have
\usepackage{jcappub} % for details on the use of the package, please
\usepackage[T1]{fontenc} % if needed
\usepackage{soul} % for using \st
\usepackage{lineno}
\usepackage{comment} 

\title{\boldmath Exploring the MeV Sky with a Combined Coded Mask and Compton Telescope:\\The Galactic Explorer with a Coded Aperture Mask Compton Telescope (GECCO)}

\author[a,b,1]{Elena Orlando}
\affiliation[a]{University of Trieste and INFN, Trieste, via Valerio 2, I-34127 Trieste, Italy}
\affiliation[b]{Hansen Experimental Physics Laboratory and Kavli Institute for Particle Astrophysics and Cosmology, Stanford University, Stanford, CA 94305, USA}
\emailAdd{orlandele@gmail.com}
\author[c,d,2]{Eugenio Bottacini}
\emailAdd{eugenio.bottacini@unipd.it}
\affiliation[c]{Dipartimento di Fisica e Astronomia G. Galilei, Univerist\`a di Padova, Padova, Italy}
\affiliation[d]{Eureka Scientific, 2452 Delmer Street Suite 100, Oakland, CA 94602-3017, USA}
\author[y,f,g]{A.A. Moiseev}
\affiliation[y]{University of Maryland at College Park, College Park, MD 20742, USA}
\affiliation[e]{University of Maryland, Baltimore County, Baltimore, MD 21250, USA}
\affiliation[f]{NASA Goddard Space Flight Center, Greenbelt, MD 20771, USA}
\affiliation[g]{Center for Research and Exploration in Space Science and Technology, NASA/GSFC, Greenbelt, MD 20771, USA}
\author[h]{Arash Bodaghee}
\affiliation[h]{Dept. of Chemistry, Physics and Astronomy, Georgia College and State University, Milledgeville, GA 31061, USA}
\author[i]{Werner Collmar} 
\affiliation[i]{Max-Planck-Institut f\"{u}r extraterrestrische Physik, Postfach 1312, 85741 Garching, Germany}
\author[j]{Torsten Ensslin} 
\affiliation[j]{Max Planck Institute for Astrophysics, Karl-Schwarzschild-Str. 1, 85748 Garching, Germany}
\author[b]{Igor V. Moskalenko}
\author[e,f,g]{Michela Negro}
\author[k]{Stefano Profumo} 
\affiliation[k]{Department of Physics, University of California, Santa Cruz, and Santa Cruz Institute for Particle Physics, 1156 High Street, Santa Cruz, CA 95064, USA}
\author[n]{Seth Digel} 
\affiliation[n]{SLAC, 2575 Sand Hill Road, Menlo Park, CA 94025} 
\author[f]{David J. Thompson}
\author[l]{Matthew G. Baring} 
\affiliation[l]{Rice University, Houston, TX 77005, USA} 
\author[m]{Aleksey Bolotnikov} 
\affiliation[m]{Brookhaven National Laboratory, Upton, NY11973,USA} 
\author[e,f,g,]{Nicholas Cannady}
\author[m]{Gabriella A. Carini} %brookhaven
\author[j,z]{Vincent Eberle}
\affiliation[z]{Ludwig-Maximilians-Universit\"at M\"unchen (LMU), Geschwister-Scholl-Platz 1, 80539 M\"unchen, Germany}
\author[o]{Isabelle A. Grenier} 
\affiliation[o]{Universit\'{e} de Paris and Universit\'{e} Paris Saclay, CEA, CNRS, AIM, F-91190 Gif-sur-Yvette, France} 
\author[p]{Alice K. Harding} 
\affiliation[p]{Los Alamos National Laboratory, Los Alamos, NM 87545} 
\author[q]{Dieter Hartmann} 
\affiliation[q]{Department of Physics and Astronomy, Clemson University, Clemson, SC 29634-0978, USA} 
\author[m]{Sven Herrmann} 
\author[w]{Matthew Kerr} 
\author[r]{Roman Krivonos} 
\affiliation[r]{Space Research Institute (IKI), Profsoyuznaya 84/32, Moscow 117997, Russia} 
\author[s]{Philippe Laurent} 
\affiliation[s]{CEA/DRF/IRFU/DAP, CEA Saclay, 91191 Gif sur Yvette Cedex, France}
\author[a]{Francesco Longo} 
\author[t]{Aldo Morselli} 
\affiliation[t]{INFN Roma Tor Vergata}
\author[w]{Bernard Philips}
\affiliation[w]{Space Science Division, U.S. Naval Research Laboratory, Washington, DC 20375, USA}
\author[y,f,g]{Makoto Sasaki} 
\author[y]{Peter Shawhan} 
\author[bb]{Daniel Shy}
\affiliation[bb]{National Research Council Research Associate at the U.S. Naval Research Laboratory, Washington, DC 20375, USA}
\author[u]{Gerry Skinner} 
\affiliation[u]{Honorary Research Fellow, School of Physics and Astronomy, University of Birmingham, UK} 
\author[y]{Lucas D. Smith} 
\author[f]{Floyd W. Stecker} 
\author[i]{Andrew Strong} 
\author[e,f,g]{Steven Sturner} 

\author[v]{John A. Tomsick} 
\affiliation[v]{Space Sciences Laboratory, University of California, Berkeley, CA 94720-7450, USA} 
\author[e,f,g]{Zorawar Wadiasingh}
\author[w]{Richard S. Woolf}
\author[y]{Eric Yates}
\author[aa] {K.P. Ziock}
\affiliation[aa]{Oak Ridge National Laboratory, Oak Ridge, TN 37830}
\author[v]{Andreas Zoglauer}

\abstract{The sky at MeV energies is currently poorly explored. Here we present an innovative mission concept that builds on and improves past and currently proposed missions at such energies. We outline the motivations for combining a coded mask and a Compton telescope and we define the scientific goals of such a mission. \\
The Galactic Explorer with a Coded Aperture Mask Compton Telescope (GECCO) is a novel concept for a next-generation telescope covering  hard X-ray and soft gamma-ray energies. 
The potential and importance of this approach that bridges the observational gap in the MeV energy range are presented. 
With the unprecedented angular resolution of the coded mask telescope combined with the sensitive Compton telescope, a mission such as GECCO can disentangle the discrete sources from the truly diffuse emission. Individual Galactic and extragalactic sources are detected. This also allows to understand the gamma-ray Galactic center excess and the Fermi Bubbles, and to trace the low-energy cosmic rays, and their propagation in the Galaxy. 
Nuclear and annihilation lines are spatially and spectrally resolved from the continuum emission and from sources, addressing the role of low-energy cosmic rays in star formation and galaxy evolution, the origin of the 511 keV positron line, fundamental physics, and the chemical enrichment in the Galaxy. Such an instrument also detects explosive transient gamma-ray sources, which enable identifying and studying the astrophysical objects that produce gravitational waves and neutrinos in a multi-messenger context. By looking at a poorly explored energy band it also allows discoveries of new astrophysical phenomena.
}

\begin{document}
\maketitle
\flushbottom

\section{Introduction}
\label{sec:intro}

At hard X-ray energies the sky has been observed by the coded mask instruments on board the INTErnational Gamma-Ray Astrophysics Laboratory (INTEGRAL) \cite{Winkler2003} for more than 15 years. On the contrary the sky at MeV energies currently remains poorly explored. Indeed, since the era of the Imaging Compton Telescope (COMPTEL) \cite{Comptel} on board  the Compton Gamma Ray Observatory, operating from 1991 to 2000, the sky above a few MeV  has been almost unexplored. As a consequence, at MeV energies there is a huge observational gap between X-rays and gamma rays. Many MeV Compton missions have been proposed in recent years (e.g., MEGA \cite{mega}, GRIPS \cite{grips}, AMEGO \cite{amego},  e-Astrogam \cite{eastrogam}, AMEGO-X \cite{amegox}, but none has been definitively planned to operate, except for COSI \cite{tomsick,zoglauer} that has been selected to fly in 2025. The many proposed missions show the strong interest of the scientific community on the potential return and acknowledge the importance of observing in this energy band. Indeed, the science drivers of the cited proposed missions span Galactic sources, extragalactic objects, transients, dark matter, cosmic rays (CRs), diffuse continuum emission, and nucleosynthesis of elements. A burst on these discoveries will be possible thanks to innovative capabilities of the described mission concept, which is unique.\\  
In this work we outline the scientific opportunities for studies in the MeV energy range with the mission concept for a mid-size Galactic Explorer with a Coded Aperture Mask Compton Telescope (GECCO) or with a GECCO mission. While the above mentioned missions are mainly focused on a Compton or a pair telescope, a GECCO mission takes advantage and optimizes on the Compton regime with a Compton telescope in combination with the photoelectric regime with the coded mask. With respect to other mission concepts, the lower energies reached by a GECCO mission and the use of a coded mask together with a Compton telescope are the keys to disentangle point sources by the real diffuse emission.

Among the most recently recognized science drivers, sensitive observations of the sky at MeV energies with unprecedented high resolution with the capability of disentangling sources from diffuse emission can open a new window to understand complicated regions such as the Galactic center, the origin of the Fermi Bubbles, the origin of the 511 keV line, possible Galactic winds, the role of low-energy cosmic rays in Galactic evolution, and their sources. It also support multimessenger astrophysics by observing and precisely localizing transients. \\   
We  briefly introduce the GECCO mission in Section \ref{sec:GECCO}, while in the following sections we  discuss the possible analysis methods to disentangle the sources from the diffuse emission. Then, we present the specific science topics that a GECCO mission will be able to address. 

\section{\label{sec:GECCO}GECCO mission}
The science objectives of the GECCO mission define the requirements for the instrument: hard X-ray - soft gamma-ray energy range, high-sensitivity, large field-of-view (FoV), and high angular ($\sim$ arcmin) and energy (order of 1\%) resolutions. It would be extremely challenging to realize all these requirements in one instrument, but a concept of several instruments may not fit the cost envelope of the mid-size mission. 

\subsection{Instrument motivation and approach}

\begin{figure*}[h]
    \centering
  \includegraphics[scale=0.4]{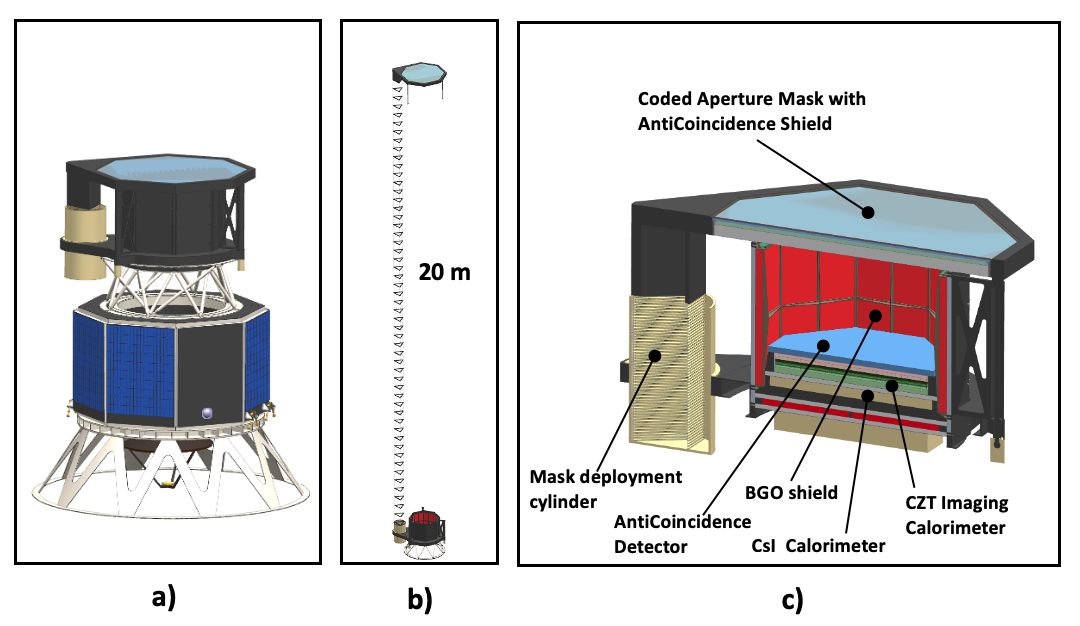}
    \caption{GECCO conceptual design: a) GECCO with Mask in stowed position and notional spacecraft bus, b) GECCO with Mask in deployed position, c) GECCO, cutaway.}
    \label{fig:GECCO_concept}
\end{figure*}

The dominating process of photon interactions with matter in the energy range from 200 keV to 10 MeV is Compton scattering, and photon detection using the Compton effect is a well established observation method in both ground-based and space-borne experiments. Compton telescopes can provide relatively low-noise observations of the large-scale diffuse radiation with a wide FoV, but their angular resolution is limited to about 0.5-3 degrees, depending on the scattering material and incident photon energy, due to Doppler broadening of the scattered photon direction induced by the velocity of the electron where the Compton scatter occurred. This is a fundamental limit, and arcmin angular resolution is impossible to achieve in a Compton telescope alone. Conversely, coded aperture telescopes are probably the only feasible way to reach arcmin and better resolution in the MeV energy range for precise localization of the point sources, but they have limited background rejection and a narrower  FoV. \textbf{The combination of a coded-aperture mask (CAM) with a Compton telescope} has been demonstrated in simulations \cite{Galloway, Aprile}, and tested with INTEGRAL/IBIS data \cite{Forot}, but the mature concept has never been implemented as the central concept for a telescope design.
This concept \textbf{will be implemented in GECCO and represents its first distinguishing feature;} it will dramatically increase the scope of the instrument and will enable the realization of the GECCO science objectives. In such an approach, the Compton telescope serves as a focal-plane detector for the CAM imaging. 

For coded-mask telescopes, the fundamental angular resolution is determined by the ratio of the CAM pixel size to the distance between the mask and the detector. Improvement of the angular resolution by reducing the mask pixel size is constrained by the available position resolution of the focal-plane detector. This is because the system's signal-to-noise ratio (SNR) strongly depends on the ratio between the detector's position resolution and the mask pixel size. Reducing this ratio improves SNR but has the opposite effect for angular resolution. For GECCO, the ratio between detector position resolution and mask pixel size was chosen to be around 0.5, which is a compromise between the opposing optimal ratios for sensitivity and angular resolution, assuming a given detector resolution.
 The other option to improve angular resolution is to  increase the distance between the CAM and the focal plane detector; however this distance is constrained by available space, in turn usually limited by the launcher shroud dimensions. An attractive option to increase the distance between the CAM and the detector is to deploy the CAM after reaching orbit, and in GECCO the CAM is deployed to 20 meters by the mast, borrowing the mast design approach from NuSTAR \cite{NuSTAR}. However, in this configuration the instrument aperture will be exposed to side-entering background from diffuse and point gamma-ray sources and from charged particles, which can significantly decrease the signal-to-noise ratio, and consequently the instrument sensitivity. Usually, in a CAM telescope, e.g., in IBIS \cite{IBIS} and SPI \cite{SPI} onboard INTEGRAL, the uncoded instrument FoV is shielded by either active thick detectors, or passive thick absorbers.  In GECCO  the problem of suppressing the side-entering background is solved by selecting the events whose Compton-reconstructed direction points to the CAM location (hereafter called Compton pointing for simplicity). \textbf{A deployed coded mask and use of Compton pointing for the background suppression is the second distinguishing feature of GECCO}, which allows greatly improved angular resolution while maintaining a high signal-to-noise ratio.

In the GECCO concept the same data set will be analyzed in two ways pursuing different science objectives. The first analysis approach uses only data from the Compton telescope to conduct  measurements with a wide FoV and modest angular resolution, enabling the sky monitoring and measurement of diffuse radiation and nuclear lines. The second analysis approach provides detection and high-accuracy localization of point sources with relatively small FoV, using CAM imaging and applying Compton pointing. As a result of this combined analysis, GECCO will create a map of all detectable sources in the IC Compton telescope $60^\circ \times 60^\circ$  FoV with modest angular resolution, with finely localized sources in the $4^\circ \times 4^\circ$ CAM FoV in the center of the Compton telescope FoV.

\subsection{Instrument design and components} 

\begin{figure}[h]
    \centering
    \includegraphics[scale=0.45]{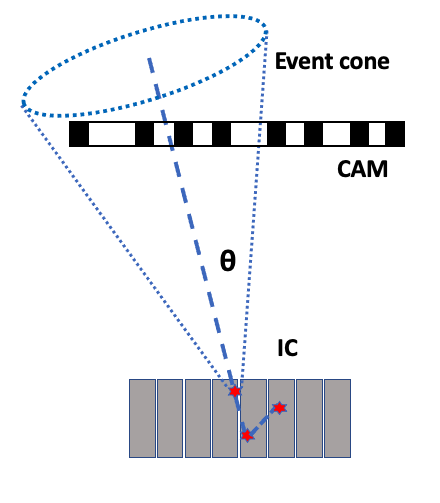}
    \caption{Illustration of the CZT Imaging Calorimeter dual capability. Red stars show the points of photon interactions detected in the IC, which are used to reconstruct the  cone of possible incident photon directions, enabling a Compton telescope functionality.
   The point of the first photon interaction, determined by the Compton event reconstruction, is used to create the CAM image, enabling the IC operation as a focal-plane detector. Dashed line shows the scattered photon direction detected by the IC, which is the axis for the Compton scatter cone.
   The dotted lines show the event cone with opening angle $\Theta$  determined by the Compton formula.}
%    \caption{Diagram of photon detection by the CZT Imaging Calorimeter. Red stars show the points of photon interactions in the CZT detector, which serves as a standalone Compton telescope and as a focal plane detector.}
    \label{fig:GECCO_event}
\end{figure}

GECCO is octagonal with a circumdiameter of $\sim$ 90cm (Fig. ~\ref{fig:GECCO_concept}). 
Such a shape provides better operation of the coded-mask instrument when compared to a rectangular shape. 
The instrument is based on a novel cadmium zink telluride (CZT) Imaging Calorimeter (IC) and a deployable CAM. It also utilizes a bismuth germanium oxide heavy-scintillator (BGO) shield, a caesium iodide  (CsI) calorimeter, and a plastic scintillator anticoincidence detector. The IC is the heart of GECCO: it operates as a Compton telescope and  serves as a focal plane detector for the CAM (Fig. \ref{fig:GECCO_event}). Its energy and position resolutions define the Compton telescope performance, while its position resolution defines the CAM pixel size and consequently the GECCO angular resolution for the CAM data analysis.

\begin{figure}[h]
    \centering
    \includegraphics[scale=0.5]{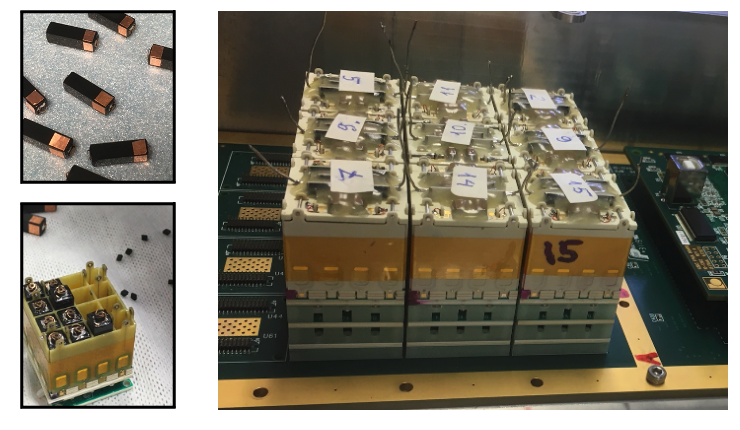}
    \caption{The components of the CZT Imaging Calorimeter. Upper left – individual CZT bars with sensitive pads, bottom left – the bars being inserted in the crate, right – Calorimeter prototype 3x3 crate array, 10cm x 10cm footprint.}
    \label{fig:Fig_CZT}
\end{figure}

\textbf{The CZT Imaging Calorimeter} detects incident photons in the energy range from  $\sim$100 keV to  $\sim$10 MeV with $>50\%$ efficiency, while measuring points of photon interaction with 3D accuracy better than 1mm and deposited energy with 1-2\% FWHM (full width half maximum) resolution above 1 MeV. The calorimeter is an array of rectangularly shaped position-sensitive virtual Frisch grid CZT detectors (bars) with baseline dimensions 8mm x 8mm x 32mm (Fig. \ref{fig:Fig_CZT}). 

The main distinctive feature of the bar detector is four 5-mm wide charge-sensing pads attached to each of its sides near the anode. The pads ensure virtual Frisch-grid effect for proper bar operation as a gamma-ray spectrometer. The signals induced on the pads, the anode and the cathode, are read out with the IDEAS-provided wave-front sampling front-end application-specific integrated circuit (ASIC) \cite{IDEAS} and used to evaluate the positions of interaction points.  The collected charge signals from the anode and the induced signals on the pads and the cathode are read out to provide X and Y coordinates by combining their ratios, while the
Z location is determined independently from the cathode to anode signal ratio and the charge drift time. In other words, each bar operates as a mini Time Projection Chamber
 (see \cite{CZT} and references therein for the detailed description of this detector). 

The bars are integrated in a 16-bar module (crate), read out by a wave-front sampling ASIC attached directly to each crate. Using this modular approach, the crates can be arranged in a Calorimeter of practically any shape and size by plugging into a motherboard, making it usable for a wide range of instruments.  A notable feature of this design is that the bars are placed “vertically”, making the effective detector thickness equal to the long dimension of the CZT bars (32mm for the GECCO baseline design). This doubles the detection efficiency achievable with the thickest commercially available CZT detectors (15mm). 

Detected points of photon interactions in the CZT bars are used to reconstruct the event cone of incident photons, enabling the Compton telescope feature. High position ($\sim$1mm) and energy ($\sim$1\%) resolutions of the CZT calorimeter are decisive in providing a reasonable Angular Resolution Measure (ARM) of 4$^\circ-$8$^\circ$, which has been proven by simulations. The ARM can be further improved by selecting events with larger distances between the first two interactions, or by checkered positioning of the bars in the crate to increase the distance between the interactions, currently being developed for GECCO. 

The MEGAlib Compton analysis toolkit \cite{MEGAlib} is used for Compton events simulation and reconstruction.  The same analysis identifies the coordinates of the photon first interaction point, which along with its measured energy enables focal-plane detector capability for the coded-aperture mask.

\textbf{The CsI Calorimeter} is positioned below the IC  and in the GECCO's baseline design is made of 4 layers of alternating orthogonal 30cm-long, 15mm x 15mm cross-section CsI logs, viewed by Silicon Photo-multipliers (SiPMs) from both ends. The energy deposited in each log is measured, and the center of gravity of energy deposition in each log is determined from the signal ratio from both log ends. The CsI Calorimeter detects energy escaping from the IC and measures the position of that energy deposition, improving the Compton reconstruction efficiency. The design of this Calorimeter is largely inherited from Fermi-LAT \cite{Atwood}.

All sides and the bottom of the CZT and CsI Calorimeters are shielded by 4-cm thick \textbf{BGO scintillator panels} which efficiently absorb the natural and artificial (produced in the instrument structure or in the spacecraft) background photons. The BGO panels also serve as a powerful quick-response GRB detector with a few degrees accuracy for GRB localization.

\textbf{Coded Aperture Mask}. Spatial modulation of the incident flux and deconvolution of the measurements from a segmented detector at the detector plane is an established method for imaging with fine angular resolution, and usage of coded-aperture masks is widespread in X-ray instruments \cite{Caroli, Skinner_basics}. A mask is an array of opaque and transparent elements set between the source field and the focal plane detector, where the latter provides the position of the detected photon interaction point and its energy. Every source within the FoV projects a shadow image of the mask onto the detector.
%(Fig.\ref{fig:Coded Mask}a). 
Techniques widely discussed in the literature allow the reconstruction of the image scene knowing the distribution and geometry of the mask pixels. They are often based on cross-correlation which can be performed efficiently using Fourier Transforms. 
The octagonal coded-aperture mask for GECCO has a circumdiameter 150 cm, which is  approximately twice as large as the IC to increase the fully-coded FoV. It is made of randomly distributed 20mm thick, 3mm square Tungsten elements.

The mask is covered by a plastic scintillator detector to veto secondary  photons which can be created by cosmic rays in the mask material. Another thin plastic scintillator detector is placed on top of the IC to veto charged cosmic rays, which otherwise would constitute a dominating background in the measurements.\\

\subsection{GECCO performance}

\begin{figure}[h]
    \centering
    \includegraphics[scale=0.55]{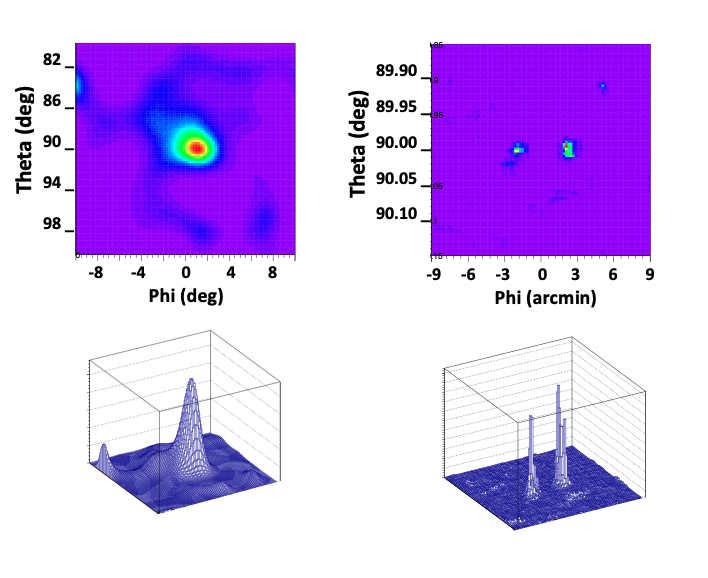}
    \caption{Simulations of the point source detection by GECCO. Left panel - Compton telescope analysis of a single source with $60^\circ \times 60^\circ$ FoV. Right panel - detection of two sources separated by 4' with the CAM data analysis. No background is included in these simulations.}
    \label{fig:Mask_images}
\end{figure}

Simulated observations of a single-source by the IC Compton telescope with the wide FoV and another of two sources separated by 4 arcmin with the CAM analysis are illustrated in Fig.~\ref{fig:Mask_images}. The effect of side-entering background in the GECCO  deployed-mask concept is being extensively studied by simulations, and preliminary results confirm the efficiency of using Compton pointing for background reduction.

The GECCO CAM data analysis utilizes the photons for which the Compton-reconstructed event ring intersects the mask location, and consequently 
the low-energy limit of this analysis will be 200-250 keV due to the prevalence at lower energies of  photoelectric absorption, for which no Compton pointing information is available.
However, it is important to lower the energy limit as much as possible due to interesting astrophysical phenomena, e.g.,  magnetar spectra. To extend GECCO acceptance to $\sim 100$ keV, defined mainly by the amount of absorbing material in the FoV (while the IC detection threshold itself can be lowered to $\sim 50$ keV), we will use the "classical" coded-mask analysis, where only a single photon interaction is needed. There will be side-entering background, causing GECCO sensitivity to degrade, but it will still be reasonably good.

The sensitivity of MeV-range instruments is strongly affected by various backgrounds of different nature, especially by nuclear activation, which is hard to predict and suppress but is very pronounced.
  These backgrounds include bright albedo and Earth limb radiation, Galactic diffuse radiation, background nuclear lines from the instrument and spacecraft, and nuclear lines produced by activation of the instrument and spacecraft by charged cosmic rays. The experience from SPI and IBIS onboard INTEGRAL pointed to the last component as especially dangerous and very hard to fight because this radiation usually is delayed after activation occurs and therefore cannot be removed by  anti-coincidence detectors \cite{Lebrun, Weid, Segreto, Sturner}. A Low-Earth equatorial orbit is currently chosen for the GECCO mission to minimize the time spent by the spacecraft in the South-Atlantic Anomaly (SAA) region that has the very high fluxes of trapped charged particles, that cause most of the activation. Also, such an orbit has the highest orbit-average vertical geomagnetic cutoff of 11-12 GV which prevents the higher fluxes of lower-energy charged cosmic rays from reaching the instrument and causing additional activation. At the instrument design level, background suppression in GECCO is implemented by placing all GECCO detectors inside a thick active BGO shield, by designing the mechanical structure with predominant use of composite (non-metal) materials to minimize activation, and by covering the CAM by a highly-efficient plastic scintillator to veto background secondary photons produce in the CAM by incident charged cosmic rays. 
  
\begin{figure}[h]
    \centering
    \includegraphics[scale=0.35]{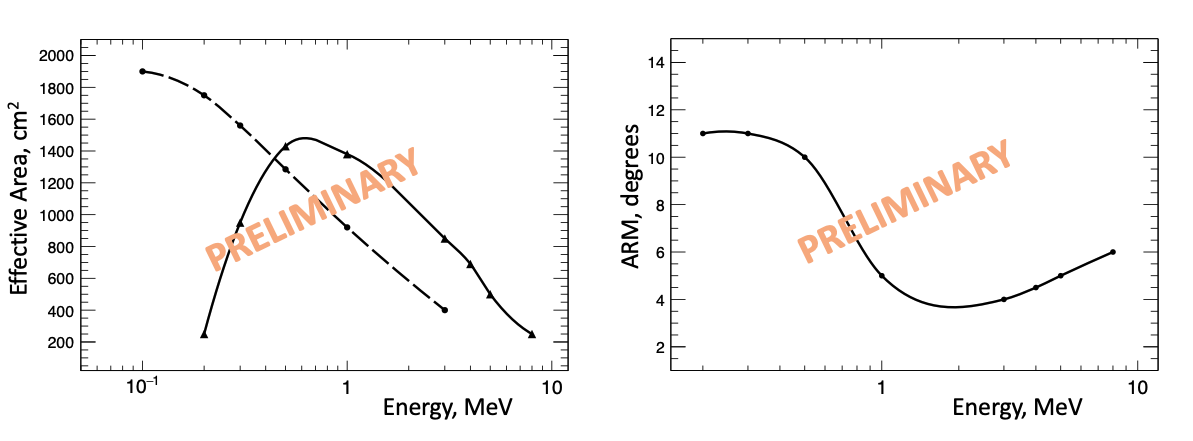}
    \caption{Simulated GECCO performance vs. incident photon energy. Left panel - effective area for coded-mask imaging, the solid line is for Compton pointing used, and the dashed line is for "classical" mask analysis. Right panel - ARM (angular resolution measure) for the IC Compton telescope.}
    \label{fig:GECCO_performance}
\end{figure}

\begin{figure}[h]
    \centering
    \includegraphics[scale=0.45]{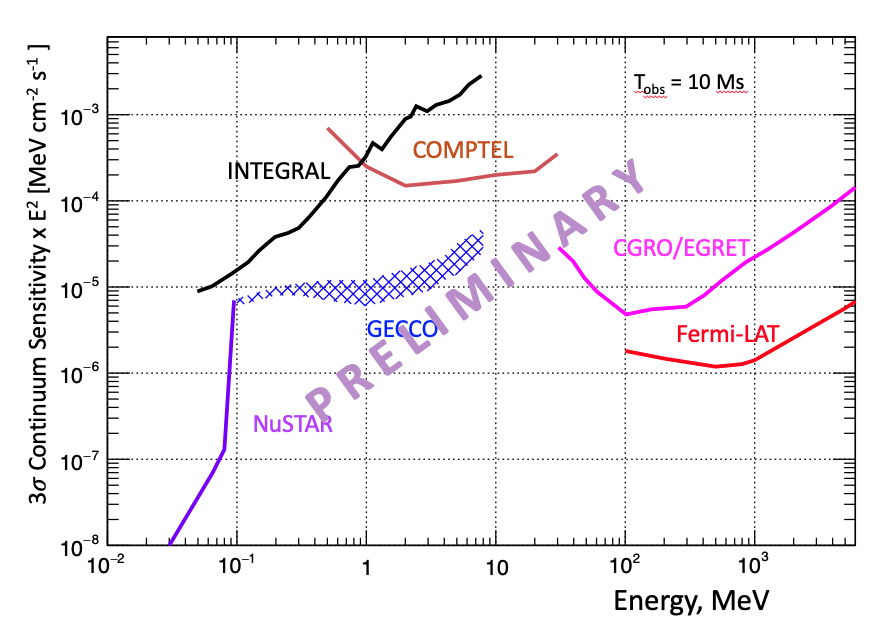}
    \caption{3$\sigma \ 10^6 s$ GECCO continuum sensitivity ($\Delta E=E$), compared with other missions. Shaded area reflects the calculation and assumptions uncertainty.}
    \label{fig:GECCO_sensitivity}
\end{figure}

  Our current estimate of $ 3\sigma$ continuum sensitivity in the observations with the CAM, based on GECCO baseline performance simulated with MEGAlib (Fig.~\ref{fig:GECCO_performance}), is shown in Fig.~\ref{fig:GECCO_sensitivity} along with performance of  current and past missions. The major factor in this estimate was to make a plausible assessment of the background reduction by the Compton pointing method, in which the solid angle of the background acceptance is reduced to the solid angle of the event cone.
  Currently, we are running massive and complicated simulations of GECCO performance, and our presented calculated sensitivity assessment is consistent with the preliminary results of these simulations. Assuming the extensive work on the analysis improvement continues, in particular in the Compton reconstruction efficiency and accuracy including use of the neural network method \cite{Zoglauer}, we have good grounds to expect further improvements on this sensitivity.

\textbf{The expected performance} of GECCO is the following: operation in the 100 keV$-$10 MeV energy range, with energy resolution of $<1\%$ from 0.5$-$5 MeV. In the CAM data analysis the angular resolution is  $\sim$1 arcmin with a $4^\circ \times 4^\circ$ fully-coded FoV, while in the Compton telescope analysis the angular resolution is 4$^\circ-$8$^\circ$ with a $60^\circ \times 60^\circ$  FoV. The $3\sigma, 10^6 s$ sensitivity is expected to be about 10$^{-5}$ MeV cm$^{-2}$ s$^{-1}$ over the entire energy range (Fig. ~\ref{fig:GECCO_sensitivity}).
GECCO can operate in either scanning or pointed mode. In scanning mode, it will mainly observe the Galactic Plane. It will change to pointed mode to either increase observation time for special regions of interest, (e.g. the Galactic Center) or to observe transient events such as flares of various origins or gamma-ray bursts. \\

\section{Point sources and diffuse: coded-mask versus Compton}

\subsection{Coded-mask mode and the INTEGRAL heritage}
For coded-mask imaging systems an astrophysical source illuminates the coded mask that casts a shadowgram onto the pixel detector. Ideally, this shadowgram is unique allowing for the reconstruction of the incidence direction of each source on the sky. This implies two fundamental requirements for coded-mask telescopes: 1) the geometric arrangement of the mask must be such that for different incidence directions the shadowgram can be uniquely identified; 2) the detector plane must be position-sensitive to actually be able to register a shadowgram. Such an imaging technology has been successfully used by instruments on board the GRANAT, BeppoSAX, INTEGRAL, and Swift missions. Basic introductions to this imaging technique can be found in \cite{caroli87} and in \cite{skinner95}.
Unlike in a conventional imaging systems, in which the recorded image is readily apparent due to the photon counts in the pixel detector, in coded-mask systems the sky image S is encoded through the mask M (a matrix of opaque and transparent elements to the radiation) in the pixel detector D. Very importantly the latter contains also an unmodulated background term B, which is due to the large collecting area. Since this latter term largely affects the noise, it enters the determination of the detection significance (signal-to-noise ratio) of astrophysical sources for background-dominated instruments. For a more precise matrix notation:
\begin{equation}
D = M~\circledast~S + B
\end{equation}

where the convolution ($\circledast$) of two generic matrices $X$ and $Y$ can be written as:

\begin{equation}
(X~\circledast~Y)_{i,j}=~\sum_{k} \sum_{l}X_{k,l}~Y_{(i+k),(j+l)}
\end{equation}

To reconstruct the sky image S$'$ a decoding function G is needed such that:
\begin{equation}
S' = G~\circledast~D 
=G~\circledast~(M~\circledast~S) + G~\circledast~B
\end{equation}
Ideally, for a perfect imaging system S$'$ = S. Therefore, according to the equation above the decoding function G must be such that (G~$\circledast$~M)~=~$\delta$-function and simultaneously (G~$\circledast$~B)~$\simeq$~0, which in actual practice is difficult to achieve \citep{skinner94}. To improve the deconvolution results, very accurate ground-based and in-flight background modeling is needed. The in-flight background modeling will be an important task for a GECCO mission. Very specifically, the open configuration of the telescope and the ability to reconstruct the Compton events allow for accounting for the astrophysical background. A more detailed discussion on this ability can be found in sections 3.3.1. and 3.3.2.

Crucial to the performance of a coded-mask imaging system is the significance at which an astrophysical source can be detected above the background. Given that roughly half of the incident photons from astrophysical sources are blocked by the mask, the detection of sources is in any case more difficult than without the mask. Yet, the actors at play are rather well defined. Thus, the significance depends on the decoding (shown above), on the open fraction $\rho$ of the mask pattern, on the astrophysical background $B$ and the detector background $b$, on the intensity of the source $S_{1}$ (that is being considered for detection), and on the remaining number $n$ of astrophysical sources $S_{i}$ in the field of view as they illuminate the detector plane. The flux contribution of these sources acts as a background term for the source $S_{1}$ which is to be detected. Therefore, it is important to account for the contribution of these sources, especially in crowded sky areas (e.g. Galactic plane) where several sources can be found in the detector's field of view. These sources can also be variable. The contribution by these sources to the overall background can be accounted for by iteratively subtracting the modeled shadowgram of each source $S_{i\neq1}$ in the field of view, which is cast onto the detector as shown for Swift/BAT and INTEGRAL/IBIS \citep{bottacini12}. The subtraction will be performed in the detector space before the decoding. This allows also to naturally account for the variability of the sources when mosaicking observations for monitoring or survey purposes. The signal-to-noise ratio \(\frac{S}{N}\) for a source with a $\delta$-function PSF is given by \citep{zand94}:\\

\begin{equation}
\frac{S}{N}~=~\frac{S_{1}}{\sqrt{\frac{S_{1}+b}{\rho}+\mathcal{B}}}
\end{equation}
where
\begin{equation}
\mathcal{B}=\frac{B+\sum_{i\neq1}^{^n}S_{i} + b}{1-\rho}
\end{equation}

As a final step the detected source $S_{1}$ can be treated as a source in the field of view and the entire analysis can be rerun to detect possible fainter sources.

\subsection{Compton mode and the COMPTEL heritage}
The Compton telescope COMPTEL (1991-2000) on the Compton Gamma-Ray Observatory (CGRO) was the first and up to now the only
Compton telescope in space. It covered the energy range 0.75 to 30 MeV, a region hardly explored in astrophysics.
Because no successor is in space yet,
the COMPTEL data are still the main astrophysical resources in this MeV gamma-ray range.

%\subsubsection{Instrument and data analysis}}
COMPTEL was a double Compton-scatter telescope without event tracking.
%{(Fig.~\ref{comptel_sketch}). 

It was sensitive to photons at soft MeV energies, i.e. 0.75 -- 30 MeV,
with an energy-dependent energy and angular resolution of 5~-~8~\% (FWHM) and 1.7$^{\circ}$-- 4.4$^{\circ}$ (FWHM), respectively.  
It had a large field of view of $\sim$1 sr and could detect gamma-ray sources with a positional accuracy 
of 1$^{\circ}$--2$^{\circ}$, depending on source flux \citep{schoenfelder93}.
%***********************************************************************************************
%\begin{figure}[h]
%\centering
%   \includegraphics[width=0.5\columnwidth]{comptelsketch.png} 
%\caption{
%A sketch of the COMPTEL instrument (from \cite{schoenfelder93}), showing the two layers of detectors as well as the detection principle of a photon in Compton mode.
%}
%\label{comptel_sketch}
%\end{figure}
%***********************************************************************************************
COMPTEL, being a ``first-generation'' instrument,  suffered from a high instrumental background. 

The COMPTEL data analysis is usually done in a so-called three-dimensional data space, consisting of the scattered photon
directions as x, y coordinates with the calculated scatter angle as z coordinate. These three quantities define a
cone-shape point-source response in such a data space. 

\begin{figure}[h]
\centering
   \includegraphics[width=0.6\columnwidth]{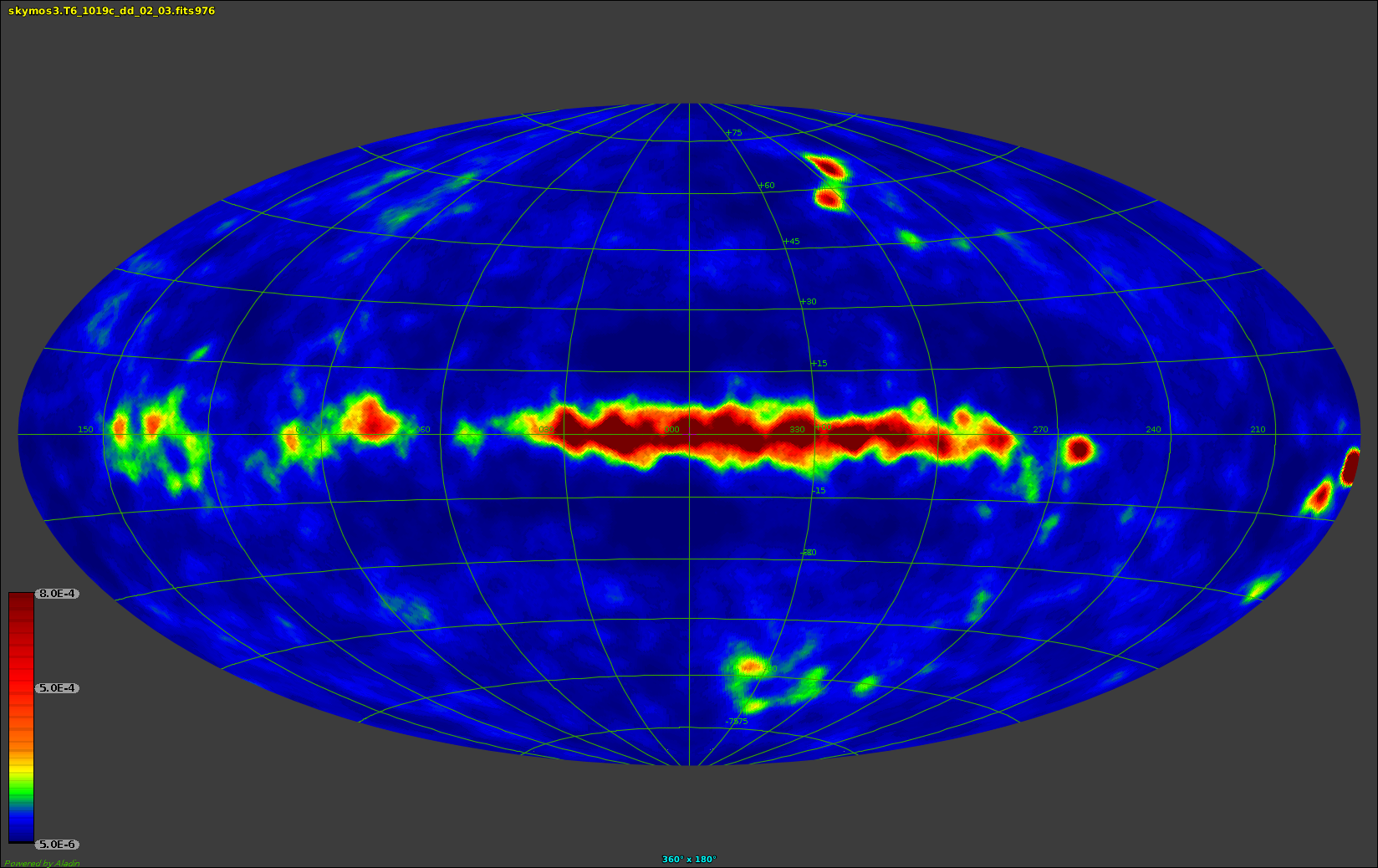}
\caption{
COMPTEL all-sky all-mission intensity map in the 9-30 MeV range, using the updated maximum-entropy method.
Evidence for several Galactic and extragalactic point sources as well as Galactic diffuse emission is clearly 
visible. 
}
\label{skymos_930}
\end{figure}

%\subsubsection{Imaging }
Imaging is challenging because only the scattered photon direction and energy deposit %in D1, D2 
are measured,
so incoming photon directions are just constrained to circles on the sky via the Compton scattering formula;
in fact these are annuli due to the measurement uncertainties. One method used with success is maximum entropy imaging (MEM)
which is in fact well suited for such problems where image and data space are quite separate \citep{strong95}.
Another imaging method used in the COMPTEL data analysis is the maximum-likelihood method (MLM) \citep{deboer92},
which is usually applied to derive source parameters like detection significances, fluxes and flux errors by
a combined model fit of a background model and various source and/or diffuse emission models. While the MEM approach, 
generating intensity maps, is superior in the overall imaging of the MeV sky, the MLM approach, generating flux
and significance maps, is superior in the quantitative analysis of point sources. Recently the MEM approach was updated for 
1) novel methods of a fast convolution-on-the-sphere and 2) the HEALPix\footnote{http://healpix.sourceforge.net} \cite{healpix} all-sky equal-area pixelization concept in 
order to generate all-sky images much faster and with finer angular resolution \citep{strong19}. An example of a
recent all-sky all-mission map in the 9-30 MeV band is shown in Fig.~\ref{skymos_930}. 
%***********************************************************************************************
%***********************************************************************************************
%\subsubsection{COMPTEL science heritage }
COMPTEL  opened the soft MeV gamma-ray band (0.75-30 MeV) as a new astronomical window, thereby bridging the gap between 
hard X-rays and medium energy gamma-rays ($>$100 MeV). The first COMPTEL source catalog \citep{schoenfelder00}, 
mainly a summary of published results of the first 5.5 years of the mission, reports 32 sources ($>$ 3$\sigma$) of various types, 
such as AGN, spin-down pulsars, gamma-ray binaries, gamma-ray line sources and extended emission regions. AGN, in particular blazars, are 
the majority of the COMPTEL point sources. Recent analyses, using data of the full COMPTEL mission and the newest analysis 
techniques, enlarge this number of point sources by typically a factor of 1.5 \citep{collmar22}. 
The Galactic diffuse emission in the COMPTEL band was studied as well \citep{strong97, strong99}, resulting for the inner
galaxy in a spectrum which is  dominated below 10 MeV by inverse-Compton emission and above 10 MeV by a combination
of inverse-Compton and bremsstrahlung emission.\\

%********************************************************************************
\subsection{Separating point sources from diffuse emission in a GECCO instrument}
%beginning of the revised version proposed to Torsten in red 

The separation of point sources from diffuse emission is a common problem in astronomical imaging, and a large number of approaches have been developed to deal with it~\cite{bertin1996sextractor, stetson1987daophot, guglielmetti2009background, popowicz2015method}. 
%\textbf{(more papers? Starblade?)}
Due to the sophisticated instrument response of GECCO it is worth thinking through this problem from the very beginning. We will see that this leads naturally to information field theory (IFT) \cite{ensslin2019information, ensslin2013information, ensslin2009information}, a probabilistic description of the problem involving field-like quantities. A good part of the  existing approaches can then be understood as different (approximate) solutions to the sky brightness field inference problem, based on a number of differing prior assumptions. 

We start this discussion with generic considerations about the separation of point sources from diffuse emission, before we discuss GECCO-specific particularities.

\subsubsection{Generic considerations}

The diffuse gamma ray flux is dominated by the emission from the Milky Way. Thanks to our position within the Galaxy, the flux reaches us from all directions, but with a clear preference for directions in the Galactic plane. Point sources can in principle appear at any sky location and in nearly any intensity. The separation of the point source and diffuse flux sky contributions therefore requires the reconstruction of two sky images, one for each of these components. Let us call them $p$ and $q$, respectively, so that the total sky flux $f=(f_x)_{x\in\mathrm{sky}}$ as a function of the sky position $x$ is $f_x=p_x+q_x$.

Even with a perfect instrument, which would map the sky brightness completely, noiselessly, and with arbitrary resolution, the separation of one observed sky brightness distribution into two is an challenging task, as 
each of those could explain the full data. The separation is nevertheless meaningful, as the idealized concepts of point sources and diffuse emission capture coarsely relevant physical concepts. Point sources are very localized compact objects and diffuse emission results from interstellar processes. 

In order to achieve such a separation the concepts of point sources and diffuse emission have to be used as discriminating criteria. Doing so may require a probabilistic or Bayesian perspective on the problem, as this provides a natural framework for incorporating prior knowledge. In this section we describe this approach.

A description of the measurement process in terms of a likelihood $\mathcal{P}(d|f)$ is necessary, incorporating the signal response consisting of point-spread and energy dispersion functions as well as the Poisson statistics of the shot noise. Here, $d$ denotes the data. In addition to this, priors for the sky brightness distributions of the two components $\mathcal{P}(p)$ and  $\mathcal{P}(q)$  are required as well.
These should encode our knowledge of the sky before measurement, but only in a generic way, so as not to determine our scientific results beyond the introduction of the two components mentioned above.

For the point source sky, a model as described below might be considered. As point sources could be anywhere, and these are largely uncorrelated (despite some preference to appear in the Galactic plane for Galactic sources) a pixelized sky map with a sufficiently high resolution should represent the point source sky, with a potential point source at each pixel location, and their fluxes being a priori uncorrelated with each other. The absence of a point source would then simply be represented by a vanishing flux at the corresponding location. 
With $p_x$ being the point source flux at pixel $x$, the prior for the point source sky would be separable into individual single point source flux priors  $\mathcal{P}(p_x)$,

\begin{equation}
    \mathcal{P}(p) = \prod_{x\in\mathrm{sky}} \mathcal{P}(p_x).\label{eq:point-s}
\end{equation}
As a priori no location should be singled out, $\mathcal{P}(p_x)$ is to be taken the same for all locations and encodes the point source brightness distribution function. This function is either postulated, e.g. a power law with high and low brightness cut offs, or better, inferred together with the point sources. For the latter option, hyper-priors that encode natural assumptions on  $\mathcal{P}(p_x)$ have to be formulated, for example that it is a strictly positive, preferentially smooth function, with a preference for power-law like slopes.  All this can easily done within the language of IFT. This point source prior, a power-law-like falling brightness function $\mathcal{P}(p_x)$ for high flux values $p_x$, can be regarded as a sparseness enforcing prior, as it will prefer that some flux within a resolution element of the instrument is represented by a single bright source over the possibility of an ensemble of dim sources, which share the observed flux in similar parts.\footnote{The reason for this is that with a power-law-like single source flux prior, the decrease in prior probability by brightening a pixel by some factor can be compensated by making a dim pixel within the same resolution element dimmer by the same factor. The total flux within the resolution element, however, increases by this operation. Thus, explaining the observed flux in a resolution element with only a single pixel strongly excited is preferred, leading to the mentioned sparseness enforcement.}

For the diffuse emission prior, a number of plausible assumptions are possible. Here, a 
minimalist choice should be discussed. Diffuse emission is characterized by exhibiting  a more or less smooth sky brightness distribution $q=(q)_{x\in\mathrm{sky}}$. This means that the sky flux does in general not change erratically from one location to the next, as the point source sky flux does, but that it is spatially correlated.  It can, however, vary largely from one area to the next, with brightness differences by orders of magnitude, but always being positive. A minimalist model (or maximum entropy model) incorporating these assumptions is that of a log-normal model, in which a Gaussian process determines the log-brightness of the diffuse sky $s=(s_x)_{x\in\mathrm{sky}}:= (\ln q_x)_{x\in\mathrm{sky}}$, with
\begin{equation}
    \mathcal{P}(s) = \mathcal{N}(s|\overline{s}, S) = \frac{1}{\sqrt{2\pi\,S}}\exp\left(\frac{1}{2}\,(s-\overline{s})^\dagger S^{-1} (s-\overline{s})\right)
\end{equation}
where $\overline{s}$ is the average log-sky brightness and $S=(S_{xy})_{x,y\in\mathrm{sky}}$ the two-point correlation structure of $s$. As both are unknown a priori, they might be inferred as well. This is possible, if we restore to the a priori assumption that no location on the sky is singled out and therefore $S_{xy}=C_s(x-y)$ should be a function only of the distance between $x$ and $y$. Then we seek only a one dimensional function $C_s(r)$ and this can be easily done with the instruments of IFT. 

This prior for diffuse flux can be regarded as a generalization for many Tikhonov regularization schemes, which are based on quadratic functionals of the regularized quantity, here $s$. It does not, however, enclose so called \textit{Maximum entropy priors}, as these can be shown to be separable w.r.t. the sky position, i.e. to be of the structure of our point source prior (Eq.~\ref{eq:point-s}), just with a very peculiar assumed luminosity function \citep[B.6]{junklewitz}. Furthermore, we note that the assumption of Gaussianity is not necessarily the only possible one.  

The (Gaussian process or other) prior for $s$ specifies $P(q) = P(s\!\!=\!\!\ln q)\, ||\partial s/\partial q||$ and these or similar assumptions specify the full Bayesian model, as the probability for all sky components and data realizations can now be specified,
\begin{equation}
    \mathcal{P}(d,p,q)= \mathcal{P}(d|f=p+q)\,\mathcal{P}(p)\,\mathcal{P}(q).
\end{equation}
From this, the posterior probability
\begin{equation}
    \mathcal{P}(p,q|d)= \frac{\mathcal{P}(d,p,q)}{\mathcal{P}(d)}
\end{equation}
allows us to make statements about the most probable sky flux distributions (the maximum a postiori estimator), their posteriori means and uncertainty dispersion. The numerical infrastructure to perform these calculations at least approximately is already in place \cite{selig2013nifty, steininger2019nifty, arras2019nifty5} and has been used to develop a point source separating imaging algorithm incorporating the above described priors~\cite{selig2015denoising}. This was even extended into the spectral domain \cite{pumpe2018denoising}, and successfully applied to data \cite{selig2015denoised, selig2015imaging}. 

\subsubsection{GECCO specific considerations}

The particularities of the GECCO instrument enter the above discussion via the likelihood function $\mathcal{P}(d|f)$.  This is key to inferring
%localize
the possible locations from which an observed photon might have come. GECCO offers two constraints on this, one via the Compton measurement and one via the coded mask. Both restrict the sky area for possible photon origins, and the more they do so, individually or jointly, the better the imaging and the separation of point sources from diffuse flux will be.

The implementation of the likelihood may require some technical developments as well. The reason for this is that the data space is six dimensional, with two photon interaction points and two energy deposition. The instrument response function is therefore a mapping from a three dimensional emission field (as a function of sky position and photon energy) into a six dimensional data space. Fast implementations of this mapping, as well as its adjoint operation, the back-projection of data space locations to possible signal space locations an observed photon could have originated from, will be required for high-performance, high-resolution imaging. These will probably be based on machine learning technologies, and exploratory studies in this direction are under way.

%**********************************************************************************
\section{Science drivers for a GECCO mission} % with a combined coded mask and Compton telescope}
\subsection{Interstellar Emission and cosmic rays}
A GECCO mission allows separation of point sources and truly diffuse emission, allowing Galactic diffuse emission and CR studies to be addressed. In fact, the larger amount  of the diffuse emission has interstellar origin. In more detail, the gamma-ray interstellar emission is produced by interactions of Galactic CRs with gas and photons as CRs propagate from their sources throughout the Galaxy.  Observations to date by the Fermi LAT, INTEGRAL, and COMPTEL underline some discrepancies with present interstellar models, leaving open questions on the large-scale distribution of CR sources, on CR transport mechanisms in the Galaxy, and on their density and spectral variation over the Galaxy (see e.g. \cite{diffuse1, Orlando18, Strong2007, Grenier, Drury} and reference therein). 
 Moreover, Galactic CRs with energies below a few GeV/nucleon and their associated gamma-ray emission are barely addressed with present telescopes. These low-energy CRs contain the majority of the energy density of the CRs. They are the main source of ionization, which affects star formation, and they  provide  pressure  gradients to  support  large-scale outflows and Galactic winds, which affect the evolution of the Galaxy. A GECCO mission assesses for the first time this low-energy CR population. In particular, for the first time it provides observations of CR electrons and positrons distributions across the Galaxy, allowing separate determination of CR leptons from hadrons. This is possible thanks to the capability of observing the inverse Compton emission component, which is related to CR electrons and the Galactic photons, after removing the source contamination. A GECCO mission also provides the first nuclear spectroscopic observation of the low-energy CRs, allowing the study for the first time of spectra, composition, and distribution of low-energy CR nuclei across the Galaxy. 
 The focus of this section is the large-scale continuum emission and the de-excitation nuclear lines.

\subsubsection{Continuum emission} 
The large-scale continuum interstellar emission in gamma rays is produced by CRs interacting with the interstellar medium, interstellar photons, and the CMB, the cosmic microwave background. 
The hadronic gas-related pion-decay emission is the major interstellar component at GeV energies, while below 100 MeV most of the emission comes from inverse-Compton scattering and from Bremsstrahlung  due  to  CR electrons \cite{Orlando18,Strong2011}. 
Observations of the large-scale Galactic gamma-ray interstellar emission from 50~keV to 10~MeV provide insights on CR sources, electron spectra, density, distribution, propagation properties, and the CR interplay with the magnetic field across the Galaxy. Indeed, below 10 MeV the continuum interstellar gamma rays are almost totally produced by low-energy CRs inverse-Compton scattering on Galactic photons (infrared, optical, and the CMB) \cite{Orlando18,Porter2008}.

A recent work \cite{Orlando18} has compared the expected interstellar emission by inverse Compton with data of the diffuse emission at X-ray and soft gamma-ray energies. Details on the inverse Compton interstellar models for that work are as follows. Propagation parameters were defined in such a way that the modeled local interstellar CR spectra and abundances reproduced the latest precise CR measurements by AMS02 \cite{AMS02} and Voyager I \cite{Voyager} after propagation. CR electrons and secondary positrons were also constrained by local gamma-ray data and especially by synchrotron data in radio and microwaves (note that the same CR electrons and positron that generate the inverse Compton emission also produce interstellar synchrotron emission by spiralling in the Galactic magnetic field). 

The CR propagation was calculated with the GALPROP code\footnote{http://galprop.stanford.edu/} \citep[e.g.][]{Moskalenko2000, Moskalenko2015, Strong2004, Gulli} accounting for the recent extension of the code to synchrotron emission and 3D models of the magnetic field \cite{SOJ2011, O2013} (for the effect of the 3D model of the magnetic field on the inverse Compton spatial distribution see \cite{orlando2019}).
Details on the data in \cite{Orlando18} are as follows. Data were taken by  
INTEGRAL \citep{Winkler2003} with its coded-mask telescope SPI, the SPectrometer for INTEGRAL \citep{SPI}. A detailed study by \cite{Bouchet} provided spectral data of the Galactic diffuse emission for energies between $\sim$80 keV and $\sim$2 MeV from  2003 to 2009 for the inner Galaxy region. 
For the same sky region intensity data at somewhat higher energies (1--30 MeV) were provided by \cite{Strong1999} from COMPTEL 
in three energy bands: 1~--~3~MeV, 3~--~10~MeV, and 10~--~30~MeV.
SPI and COMPTEL data were both cleaned by subtracting the sources \citep{Strong1999, Bouchet}.
The conclusion of \cite{Orlando18} was that the best model described above underestimates the X-ray emission in the inner Galaxy.
The same authors suggest that SPI and COMPTEL diffuse data in the inner Galaxy region may be affected by contamination from unresolved sources (due to the well-known limited sensitivity and angular resolution of the instruments). Such a possible contaminating source population in the SPI and COMPTEL energy band could be the soft gamma-ray pulsars that were found to have hard power-law spectra in the hard X-ray band and reach maximum luminosity typically in the MeV range \citep{Kuiper}.  

A GECCO mission is able to detect these potential sources and definitively disentangle the true diffuse emission from possible unresolved sources.
This also enables study of low-energy CRs that are thought to be a fundamental component of the interstellar medium, but whose composition, distribution, and flux are poorly known. 
Observations at soft gamma-ray energies and below would inform on the large-scale distribution of CR sources, on CR transport mechanisms in the Galaxy, and on their density and spectral variation over the Galaxy (see e.g. \cite{Strong2007, OrlandoWP2019}). 
 Observations at soft-gamma rays also provides information about the interplay of low-energy CRs with Galactic winds and on the role of low-energy CRs on Galaxy evolution. The connection between low-energy CRs below a few GeV/nuc and galaxy evolution has started to be investigated only recently and is still poorly understood (e.g. \cite{Ensslin, Persic, Padovani, Hanasz2009, Giacinti2012, Recchia2016, Pfrommer2017}). 
Even more specifically, a GECCO mission for the first time allows observations of the emissions from CR electrons clearly separated by the emission from CR nuclei. It will reveal the spatial and spectral distributions of the inverse Compton emission in the Galaxy \cite{Orlando18}, important for disentangling emission not only from unresolved sources (e.g. \cite{Strong}), but also from the extragalactic diffuse gamma-ray background (e.g. \cite{EGB}), or from potential signals of dark matter annihilation (e.g. \cite{FermiGC}), which have distributions similar to the inverse Compton component. Such observations also allow inferences about the distribution of CR electrons, which best sample CR inhomogeneity, because they are affected by energy losses more strongly than nuclei, and they remain much closer to their sources. Moreover, observations of gamma rays below 10 MeV produced by the same electrons that produce synchrotron emission in radio and microwaves provide firmer constraints on Galactic magnetic fields (see e.g. \cite{O2013, Orlando18, orlando2019}).

\subsubsection{De-excitation nuclear lines}

Gamma-ray lines in the 0.1 - 10 MeV range are produced by nuclear collisions of CRs with interstellar matter \cite{Bykov01}.
Their detection allows study of the spectra, composition, and distribution of CR nuclei below the kinetic energy threshold for production of neutral pions ($\sim$300 MeV for p+p collisions). The most intense lines are expected to be from the de-excitation of the first nuclear levels in $^{12}$C, $^{16}$O, $^{20}$Ne, $^{24}$Mg, $^{28}$Si, and $^{56}$Fe \cite{Ramaty}. The total nuclear line emission is also composed of broad lines produced by interaction of CR heavy ions with the H and He nuclei of the interstellar gas, and of thousands of weaker lines \cite{Bykov01}. The gamma-ray spectrum is predicted to have a characteristic bump in the range 3 - 10 MeV, which is produced by several strong lines of $^{12}$C and $^{16}$O. Simulated gamma-ray line spectra  of an individual nearby superbubble is reported in \cite{tat04, ben13}. The spectrum comprises narrow and broad $^{12}$C and $^{16}$O lines, the observation of which would constrain low energy CR composition \cite{Bykov01}. %More details can be found in \cite{Bykov01}. 
A GECCO telescope covers the energy region where these nuclear lines are expected.

\subsection{Nucleosynthesis lines} \label{nucleo}
The sites believed to produce radioisotopes observable as gamma-ray line emission are novae, core-collapse Supernovae (SN), SN type Ia,  Wolf-Rayet stars, and asymptotic giant branch stars.
Nuclear emission lines from isotopes in massive and exploding stars, such as $^{44}$Ti, $^{26}$Al, and $^{60}$Fe, allow a probe of nucleosynthesis and chemical evolution of the Galaxy.  While the above-cited radioisotopes with relatively long lifetimes produce diffuse emission that provides insights on stellar nucleosynthesis and also on the Galactic interstellar medium, the radioisotopes with shorter lifetimes, such as $^{7}$Be, $^{56}$Ni, $^{58}$Ni, provide information about the explosion and the early evolution of the remnant.  

The all-sky COMPTEL map showed the gamma-ray emission produced by the radioactive decay of $^{26}$Al \cite{ComptelLine} to be concentrated along the plane, tracing regions with massive young stars throughout the Milky Way. More recently \cite{Diehl}, the Doppler shifts of the gamma-ray energy caused by the Galactic rotation has been observed with INTEGRAL/SPI, which depends on the location of the source region within the Galaxy, and, hence can enable a census of massive stars in the Galaxy. 
Moreover, being produced in the innermost ejecta of core-collapse supernovae, $^{44}$Ti provides a direct probe of the supernova engine. Most numerical simulations of stellar core-collapse explosions require spatial asymmetry, which has been observed in Cassiopeia A with NuSTAR \cite{Grefenstette} thanks to the detailed image of $^{44}$Ti line at  around 70 keV. This provides strong evidence for the development of low-mode convective instabilities in core-collapse SNe. Even more recently, an asymmetric explosion has been revealed with the detection of the $^{44}$Ti gamma-ray emission line from SN1987A with NuSTAR \cite{Boggs}. 
Other nucleosynthesis lines in the energy range of a GECCO mission are: $^{56}$Ni and $^{57}$Co.
A GECCO telescope allows mapping of radioactive material in SN remnants, resolving the Galactic chemical evolution and sites of nucleosynthesis of elements.

\subsection{Understanding the Galactic center gamma-ray excess}

The Galactic center (GC), a favorite target for telescopes across the whole electromagnetic spectrum, provides guaranteed exciting scientific return. The GC harbors the SMBH with mass of $4\times10^6 M_\odot$ and dense populations of all types of objects including  binary and multiple systems, while its relative proximity allows many such objects to be resolved. The two huge Fermi Bubbles, each 10 kpc across, presumably emanating from the GC to the North and to the South of the Galactic plane were discovered by Fermi-LAT in gamma rays \cite{2010ApJ...724.1044S,2014ApJ...793...64A}, and are also visible in X-rays by eRosita \cite{eRosita-FB}, testifying that this is a multi-wavelength phenomenon (for more details see Section~\ref{bubbles}). The high-energy processes that involve particle acceleration and interactions reveal themselves through generation of non-thermal emission observed from radio- to gamma rays. The GC is also bright in an enigmatic positron annihilation emission that includes 511 keV line and three-photon continuum emission \cite{2020NewAR..9001548C}. 

Recent observations of the GC with Fermi-LAT reveal an excess in the energy range around 10 GeV \cite{2011PhLB..697..412H, 2016ApJ...819...44A}. The analysis made using different techniques indicates that the excess is spatially extended and concentrated around the GC. The NFW template fitted with other templates built using a GALPROP-based diffuse emission model effectively flattens the residuals leaving a burning question about the origin of the excess open.
   
Two main interpretations of the excess relate its nature to the unresolved sources that may be abundant in the inner Galaxy \cite{unresolvedGC} or to  emission due to  DM annihilation \cite{2017PhRvD..95j3005K}. Both interpretations are supported with valid arguments that have to be tested with further observations. In particular, the DM interpretation is supported by observations of the excess in CR antiprotons and with observations of the extended 400 kpc-across the gamma ray halo around the Andromeda galaxy (M31) \cite{2019ApJ...880...95K}. In both cases the excesses are observed in the same energy range \cite{2021PhRvD.103b3027K} giving strong support to the DM scenario. Meanwhile, the conventional astrophysical interpretation in terms of the weak unresolved gamma ray sources is supported with the logN-logS plots \cite{2015ApJ...812...15B,NuSTAR-GC}.

Addressing these questions requires the high-resolution observations in the energy range covered of a GECCO mission with its angular resolution that allows to resolve sources. In fact, the currently operating Fermi-LAT instrument has very limited capabilities below $\sim$500 MeV with angular resolution becoming as bad as several degrees below 100 MeV. X-ray telescopes have the angular resolution at arcsec scale; however, their operating energy is too far below the energy scale to provide relevant information. The sources that are observed with present X-ray telescopes and the processes of generation of X-ray emission may be and likely are very different from those in the MeV scale. That diminishes their capabilities to resolve this issue.

\subsection{Searches for dark matter and new physics}
A GECCO telescope offers unprecedented opportunities in the search for dark matter and new physics \cite{geccodm,Coogan:2020tuf}. Specifically, ``light'' dark matter, in the GeV or sub-GeV mass range, has come to the forefront in the present era that has been dubbed one of the ``waning of the WIMP'' \cite{Arcadi:2017kky}. The pair-annihilation or the decay of such light dark matter particles, resulting in MeV gamma rays from a number of targets, most notably the center of the Galaxy, nearby galaxies such as M31, and nearby dwarf satellites of the Milky Way, would have escaped detection with previous telescopes, but would be detectable by a GECCO telescope.

\cite{geccodm} studied in detail the potential of GECCO to discover a signal of dark matter annihilation or decay, using the state-of-the-art code {\tt Hazma} for the calculation of the gamma-ray spectrum from simplified dark matter models matched via chiral perturbation theory onto final-state hadrons \cite{hazma} (see also \cite{hazma2}). The key findings of \cite{geccodm} are that: 
\begin{enumerate}
    \item The Galactic center is the most promising target for searches for dark matter annihilation, followed by M31 and by local dSph such as Draco; 
    \item Considering individual final states, a GECCO mission improves over current constraints from Fermi-LAT, EGRET and COMPTEL by over 4 orders of magnitude for dark matter annihilating to $e^+e^-$ and by 3-4 for annihilation into $\gamma\gamma$ or $\mu^+\mu^-$ (see fig.~1 in \cite{geccodm});
    \item For dark matter decay, the largest gains will be made for $e^+e^-$ and $\gamma\gamma$, again via observations of the Galactic center;
    \item Considering a specific simplified model, \cite{geccodm} finds that for light scalar mediators (lighter than the dark matter mass) a GECCO mission probes thermal relic dark matter in a very wide range of masses, from 0.5 MeV up to a GeV, improving by up to 4 orders of magnitude current constraints;
    \item For a vector mediator, similarly, a GECCO mission outperforms current constraints by several orders of magnitude, especially in the sub-MeV dark matter mass range.
\end{enumerate}
\cite{Coogan:2020tuf} additionally studied opportunities for constraining or discovering light primordial black holes that are currently in the process of evaporating via the mechanism of Hawking radiation. Interestingly, the expression for the approximate black hole lifetime $\tau$ as a function of the hole's mass $M$,
\begin{equation}
    \tau(M)\simeq 200\tau_U\left(\frac{M}{10^{15}\ {\rm g}}\right)^3\simeq 200 \tau_U
\left(\frac{10\ {\rm MeV}}{T_H}\right)^3,
\end{equation}
where $\tau_U$ is the age of the universe, and $T_H$ the Hawking temperature of the hole, points to temperatures at evaporation {\em at most} as large as 10 MeV. Of course more energetic particles can also be radiated via thermal fluctuations, but it is clear that the expected detectable gamma-ray emission falls squarely within GECCO observing capabilities. \cite{Coogan:2020tuf} presented an accurate evaluation of the expected gamma-ray spectra from light black hole evaporation, and showed that a GECCO mission will enable the possible discovery of light primordial black holes as massive as $10^{18}$ g as dark matter candidates, significantly extending current constraints, by up to 1-2 orders of magnitude in mass.

\subsection{The Fermi Bubbles}\label{bubbles}
% \begin{center}
%   \textcolor{red}{----WORK IN PROGRESS---- }
% \end{center}
A GECCO mission is also suitable for observing the region of the Fermi Bubbles (FB).
FB are a pair of Galactic-scale structures extending, almost symmetrically, above and below the Galactic plane. Discovered in 2010 by \cite{FBdiscovery} in a search for a gamma-ray counterpart to the WMAP\footnote{Wilkinson Microwave Anisotropy Probe: \url{https://map.gsfc.nasa.gov}.} haze (see e.g. \cite{Finkbeiner}), the FB were deeply studied in 2014 by \cite{FBLAT} who performed detailed spectral and morphological analysis for $|{\rm b}|>10^\circ$: both bubbles are elliptical, extending 55$^\circ$ North-South and 45$^\circ$ East-West in diameter; they appear to have a vertical axis (perpendicular to the Galactic plane) roughly intercepting the GC; they have an almost uniform intensity, a quite hard spectrum well described by a log parabola or a power-law with exponential cutoff; their gamma-ray luminosity between 100 and 500 MeV was estimated to be  $L\gamma = (3.5-6.8)\times 10^{37}$ erg/s and leptonic inverse Compton or hadronic (plus inverse Compton from secondary leptons) models can explain the data well. Leptonic scenarios can also explain the microwave haze observations, but hadronic scenarios do not suffer from radiative losses and can thus maintain high-energy particles even if operating on much longer timescales (although particle confinement on Gyr timescales is challenging). Assuming a jet-like FB formation from the GC, the FB expansion velocity should be greater than 20,000 km/s in order to have a bubble formation time greater than the cooling time of TeV electrons (assuming both inverse Compton and synchrotron losses in a 5 $\mu$G Galactic magnetic field); this corresponds to electron acceleration time scales of roughly 500 kyr \cite{FBLAT}. A 2019 study of the low-latitude region of the FB \cite{MalyshevHerold} found greater intensities than the FB at high latitudes with a spectrum compatible with a single power law between 10 GeV and 1 TeV and, more interestingly, a centroid shifted to the west of the GC. The latter observation disfavors models attributing the origin of the FB to past AGN-like activities of the super-massive black hole in the center of our Galaxy. 

Observing a soft gamma-ray counterpart of the FB would favor a leptonic scenario in which a low-energy CR electron population produces gamma rays below $\sim$10 MeV through inverse Compton scattering on the interstellar radiation field. On the contrary, an absence of such a counterpart would favor an origin in hadronic processes for the FB in which the main process of gamma-ray production is pion decay (completely subdominant below 100 MeV with respect to inverse Compton and bremsstrahlung). Additionally, the unique capability of a GECCO mission to resolve point-like sources along the Galactic plane will help disentangle the emission from such sources and the FB low-latitude emission, providing useful insights about the origin of these large-scale features.   

Recently eROSITA \cite{eROSITA} detected a new gigantic bubble-like feature in the Southern hemisphere of our Galaxy \cite{eRBNature}, complementary to a Northern hemisphere feature already known from X-ray and radio observations\footnote{The Northern hemisphere feature is associated with the North Polar Spur observed in X-rays \cite{NPSxrays} and Loop I observed in radio \cite{radioLoopI}.}. The eROSITA bubbles (eRB) are morphologically almost spherical, extending $\approx 80^\circ$ in diameter, and they are not obviously symmetric if considering a vertical axis passing through GC. The measured intensity between 0.6 and 1 keV is not uniform, with a total luminosity (assuming a hot X-ray-emitting plasma) of $L_{X}\approx10^{39}$ erg/s, and a measured average surface brightness of $(2-4)\times10^{-15} {\rm erg/cm}^2{\rm /s/arcmin}^2$ (assuming an emission from hot plasma  with temperature kT=0.3 keV) that decreases with Galactic latitude. In \cite{eRBNature}, assuming a Mach number of the shock of 1.5, the authors estimate a characteristic expansion time to the present size of around 20 Myr ($\approx$ 40 times the FB expansion timescales for leptonic scenarios). 

\cite{eRBNature} suggests a connection between the eRB and FB, in which the latter are driving the expansion of the former and they are both associated with the same energy release in the GC region. In this scenario the FB outflow piles up and heats the surrounding interstellar gas and the outer eRB boundary represents the termination shock of this heating wave. The pressure between the FB and eRB surfaces is constant and the total thermal energies at the two boundaries reflect their volumes (hotter plasma at the outer eRB boundary). However, although some morphological similarities exist, the connection between the eRB and the FB (and even their association to the GC itself) is not straightforward. More dedicated studies and new observations are needed to better investigate the physical relation (if any) between the FB and the eRB.  Continuum observations of gamma rays between hundreds of keV and tens of MeV could be crucial to unveil the origins of the FB \cite{NegroMeVBubbles} and possible connections between FB and eRB.

From our perspective, it is not yet known whether such gigantic bubbles are truly of galactic scales originating in the GC or if they are smaller, closer features. The Andromeda galaxy (M31) is a barred spiral galaxy like our Milky Way, and the two also share similar virial masses and reasonably similar formation stories: Andromeda is approximately a twin of the Milky Way. Observations of Andromeda provide a different perspective on our own Galaxy. For this reason, the gamma-ray observation of giant bubble-like structures extending above and below Andromeda's plane \cite{M31bubbles} is an extremely interesting piece of information, pointing toward truly galactic-scale interpretation of the FB. Recently \cite{M31imaging} provided a gamma-ray imaging of M31 which gives the visual impression of bubble-like structures, limited however by the relatively poor angular resolution of the LAT at the observed energies.  A GECCO mission could also provide a soft-gamma-ray picture of our twin galaxy, providing again very valuable hints about the origin of the FB.

\subsection{The 511 keV line} 

A 511 keV line emission from positron-electron pair annihilation in the central regions of the Milky Way was discovered by balloon-borne experiments as early as 1975 (see e.g. \cite{1975ApJ...201..593H}). Further observations with space telescopes, specifically OSSE on the Compton Gamma-Ray Observatory~\cite{Skibo:1997yk} and, more recently, the SPI spectrometer~\cite{Weidenspointner:2008zz, 2003A&A...407L..55J} and the IBIS imager on board INTEGRAL~\cite{DeCesare:2005du} have significantly sharpened the observational picture of the 511 keV line. The line intensity is, overall, around $10^{-3}$ photons cm$^{-2}$ s$^{-1}$, originating from a 10$^\circ$ region around the Galactic Center. 

New physics explanations for the 511 keV emission are constrained by observations both at higher and lower energies, indicating, for instance, that the mass of a putative dark matter candidate whose annihilation could produce the observed line is bounded from above at around 3 MeV \cite{1981SvAL....7..395A, beacom2006}. Absent large-scale magnetic fields~\cite{prantzos2006}, any astrophysical source of the 511 keV line emission should additionally lie within approximately 250 pc of the annihilation sites~\cite{jean2006}, thus implying that the source distribution should quite closely resemble the actual signal distribution in the sky~\cite{churazov2005, jean2006}.

While the nature of such astrophysical sources continues to be debated, the morphology and a lower-limit on the number of sources rules out a single source (e.g. Sgr A*~\cite{1982AIPC...83..148L}) or a single injection event, such as a gamma-ray burst or a hypernova in the Galactic Center~\cite{1984AIPC..115..558L}. The signal sources must therefore be associated with a population of sources that could, or not, be resolved as individual point sources (a possibility somewhat constrained by prior observations ~\cite{2005A&A...441..513K}). Source classes that have been considered include massive stars, pulsars, including millisecond pulsars, core-collapse supernovae and SNe Ia, Wolf-Rayet stars, and low-mass X-ray binaries (LMXB), especially microquasars~\cite{Siegert:2015knp, Bandyopadhyay:2008ts}. In many instances, these astrophysical objects are also found much closer to the solar system than in the Galactic Center region. \\
The angular resolution and point-source sensitivity of a GECCO telescope make the instrument ideally suited to enable differentiation between multiple point sources and a genuinely diffuse origin for the 511 keV emission, as expected from dark matter annihilation or other exotic scenarios. Specifically, if one source class dominated the positron emission, a GECCO mission could detect nearby members of that source class. \cite{geccodm} specifically showed that GECCO sensitivity should enable the detection of any positron source responsible for a significant fraction of the 511 keV signal closer than 4 kpc. \\
Additional information on the nature of the origin of the 511 keV signal from the Galactic Center will be provided by observations of nearby systems such as the Andromeda galaxy (M31), the Triangulum galaxy (M33), nearby clusters such as Fornax and Coma, and nearby satellite dwarf galaxies such as Draco and Ursa Minor~\cite{Wolf:2009tu}. Using as a crude estimate of the predicted 511 keV signal a simple mass to distance-squared ratio, \cite{geccodm} finds that the 511 keV signal from M31 should be detectable by a GECCO mission, as should the signal from the nearby dSph Fornax and (although marginally) the Coma cluster. \cite{geccodm} predicts that  M33, and local dSph should not be bright enough at 511 keV to be detectable by GECCO. Integral/SPI already searched for a 511 keV line from Andromeda (M31), reporting an upper limit to the flux of $1\times 10^{-4}\, {\rm cm}^{-2}\, {\rm s}^{-1}$~\cite{Bandyopadhyay:2008ts}. Certain types of new physics explanations such as dark matter decay would follow a similar scaling, while others would have a more complicated, model-specific dependence.

\subsection{Sources and source populations}
Current available observations in the MeV domain have an angular resolution of several degrees. This rather poor angular resolution is due to the changing nature of the photon-matter interaction used to detect the astrophysical radiation. 
Indeed, while at several tens of MeV pair production dominates, at lower energies at a few MeV Compton scatting is the primary interaction process, which was used by COMPTEL.
 Inevitably also GECCO pure Compton mode is affected by the large angular resolution. However, the coded-mask mode allows a GECCO mission to reach a superb, for this energy domain, angular resolution of $\sim$1 arcmin. The ability to separate the flux contribution of single sources at the arcmin level also allows precise spectroscopy. This feature helps in identifying newly detected sources in a basically unexplored energy range. In fact, while COMPTEL sources are mostly associated and/or identified through variability of exceptionally bright sources, GECCO newly detected sources can be positionally and spectroscopically identified through the contiguous energy bands of the Fermi-LAT and the well known keV sky. Here we summarize the most significant and interesting source populations, both extragalactic and Galactic that can be observed by a GECCO mission. 

\subsubsection{Extragalactic source populations}
The high-energy cosmic diffuse background radiation is a useful tool to constrain the population of astrophysical sources
that are responsible for it. This background radiation at MeV energies has been measured by COMPTEL in a study by
\cite{weidenspointner2000}, who accurately accounted for the instrumental effects. This measurement ties in well with
the measurement of the diffuse X-ray background by several instruments \citep[e.g.][]{faucher20} and the diffuse gamma-ray background
measured by the Fermi-LAT \citep{ackermann15}. The extrapolation of the latter to lower energies and the extrapolation of the
former to higher energies, require a hard MeV component, which has been measured by COMPTEL. A major contribution to the low
energy part between a few hundreds of keV and a few MeV comes from blazars that are efficiently detected in hard X-ray
($>$15 keV) due to their rather hard spectra. Among the most interesting of such sources detected at hard X-rays are the
extreme synchrotron BL Lac objects \citep[e.g.][]{costamante01} and the high-redshift blazars \citep[e.g.][]{bottacini12}.
High-redshift blazars, especially Flat Spectrum Radio Quasars, are important as they are known to host supermassive black holes of the order of 10$^{9}$ M$_\odot\ $
\citep{ghisellini10}. The existence of such massive black holes in the early universe is relevant for
scenarios in which they are formed by accretion or by merger-driven evolution. In contrast, the extreme synchrotron BL Lac objects
carry  information about the composition of the jet.  The very high-energy spectral energy distribution (SED) can be explained as due to a hadronic component
in the jet \citep[e.g.][]{cerruti15}, which can account for a significant fraction of the neutrino emission. The contribution to the diffuse
high-energy hard component measured by COMPTEL calls for candidates different from blazars. While DM can contribute
to it as discussed in section 6, also point sources different from blazars are good candidates. An intriguing class of sources
are star-forming galaxies (SFG). SFG are rich in CR that undergo hadronic interactions with the interstellar medium. This process
led to the detection of some SFG in the GeV band \citep{abdollahi20}. However, the exact contribution to the diffuse background remains unsettled
\citep{owen21}. The excellent  sensitivity and angular resolution of a GECCO mission allows for detecting and pinpointing these sources,
thereby accounting for their contribution to the diffuse background radiation. A further contributing class of sources to the high-end
of the MeV diffuse emission are radio galaxies \citep{inoue11}, which have been detected in this energy range.

\subsubsection{Galactic source populations}

The Milky Way and similar galaxies host a rich diversity of objects capable of radiating in the MeV range. Many of these objects involve a neutron star (NS) or a black hole (BH) which represent the densest forms of matter in the Universe and are the final stage in the lives of massive stars. 

Around a NS, gamma-rays can be generated by thermonuclear reactions of material on the hot surface (bursters) or by extraction of magnetic or rotational energy from the NS (magnetars and pulsars, respectively). There are 239 pulsars listed in the fourth Fermi-LAT catalog \cite{2020ApJS..247...33A}. Since a GECCO mission samples the energy band below LAT's limit of 50 MeV, it will not only expand the population of young pulsars whose emission is expected to peak in the MeV range \cite{2015MNRAS.449.3827K}, it will also fill in the gaps in the spectra of pulsars between the X-ray and gamma-ray bands.

Around a NS or a BH, gamma-rays can result from the accretion of charged particles accelerated in the strong gravitational and electromagnetic fields of so-called X-ray binaries (XRBs). There are around 400 known XRBs in our Galaxy \cite{2001A&A...368.1021L, 2006A&A...455.1165L, 2019NewAR..8601546K}. Cyclotron lines have been found in the range of 10--100\,keV for 35 XRBs \cite{2019A&A...622A..61S}, but some XRBs could host magnetars \cite[$B \gtrsim 10^{14}$\,G,][]{2008ApJ...683.1031B} that would push these lines, as well as their harmonics, to hundreds of keV where they can be seen by a GECCO mission. 
 
If the NS or BH features a jet, the X-ray photons (and UV photons from the donor star) can interact with particles in the jet causing them to be upscattered via inverse Compton to GeV energies \citep[e.g.,][and references therein]{2009IJMPD..18..347B}. Thus far, GeV emission has been detected from a dozen so-called gamma-ray binaries. Most of them have a NS as the accretor while a few have a BH: the only thing they appear to have in common is that they all have a high-mass star as the donor. Their emission is expected to peak in the MeV band, which means that a GECCO mission will connect the X-ray continuum with that from the GeV band. This connection can then be used to disentangle conflicts between leptonic and hadronic emission models. In the same way, a GECCO mission will extend the tail in the hard state of BH-XRBs into the MeV domain. %Photons originating from the base of the accretion column are expected to be polarized \citep[e.g.,][]{2011Sci...332..438L} which GECCO is conveniently built to detect. 

Before a massive star turns into a NS or a BH, it goes through a supernova (SN) phase where stellar material accelerated by the sudden collapse of the core emits gamma-rays at specific energies that reveal the star's chemical composition (Section \ref{nucleo}). Prior to the SN stage, many of these massive stars are bound gravitationally to another massive star. The shock region where the stellar winds collide can also give rise to gamma-ray emission in these colliding-wind binaries \citep[CWBs: e.g., ][and references therein]{2020MNRAS.495.2205P}. In the MeV range, a GECCO mission will link the keV to GeV continuum from CWBs such as eta Car \cite{2018NatAs...2..731H} and allow us to dissociate the contributions from leptonic (inverse Compton) and hadronic (pion decay) acceleration mechanisms.

For these reasons, when a GECCO telescope observes the Milky Way's MeV-emitting populations, it will show us different stages in the life cycle of massive stars. Once both stars have collapsed into a NS or a BH, and when the pair eventually merges into a single object, the merger produces gravitational waves detectable by the LIGO and Virgo observatories. Though such signals have been extragalactic in origin so far, predictions for the merger rate depend on knowing how many members from each of the populations above are hosted by galaxies like ours \cite{2021ApJ...913L...7A}.

\subsection{Multimessenger and multifrequency synergies}

Given the transient and variable origin of multimessanger and multifrequency astrophysical
sources, the fraction of the sky being monitored at any given time is a major asset for a
space mission. In its Compton observing mode a GECCO mission will cover a large fraction of the sky
of 60$^\circ$~$\times$~60$^\circ$ in zenithal direction allowing to keep watch over flaring
phenomena like blazars and transient phenomena like Gamma-Ray Bursts (GRBs). Also, GECCO 
BGO shielding, specifically designed for background rejection with its octagonal structure
of large-size detectors of $\sim$3000 cm$^{2}$, will have the additional ability to
locate the prompt emission of GRBs within a few degrees similar to INTEGRAL \citep{kienlin01}.
The prompt emission by merging neutron stars can be effectively observed in the GECCO energy
band ~$\sim$keV--MeV. They reveal themselves as short GRBs as well as kilonovae. Such events
also provide gravitational wave (GW) signals allowing a GECCO mission to tie in with multimessanger
and multiwavength observations. Amid the prompt-emission detection, the telescope can repoint within
a few minutes depending on the slewing angle, allowing for locating the source within better
than 1 arcmin precision. It will also act as an alert system for follow-up observations. The
study of neutron star mergers provides insights into relativistic jets and particle physics.
Neutron stars might also be involved in the emission of very short GRBs when transitioning
to strange quark stars \citep{alcock86}. While this intriguing hypothesis is still an open
question, it enables studies related to fundamental physics of matter. In a multifrequency
approach, GECCO large field of view allows for the coverage of the little explored MeV
range of flaring sources. Such sources can be galactic or extragalactic in origin. Among
the extragalactic sources blazars represent a major discovery space. Indeed, a tentative 
$\sim$~3$\sigma$ association of a high-energy neutrino detected by IceCube with a flaring
blazar \citep{aarsten18} has revived the lepto-hadronic emission scenario for these sources,
which would favor the neutrino production in the jet. The energy band of ~$\sim$keV--MeV carries
the signature to constrain the content of the jet \citep[e.g.][]{gao19, reimer19, bottacini16}.

\section{Conclusions}
In this work we have presented a novel mission concept for a next-generation telescope covering  hard X-ray and soft gamma-ray energies, the GECCO Galactic Explorer with a Coded Aperture Mask Compton Telescope.
We have discussed the importance of a mission like GECCO, which combines a coded mask with a Compton telescope, that will finally cover the huge observational gap between X-rays and gamma rays.
The new mission concept of combining the high-resolution of the coded mask with the high sensitivity of the Compton telescope will allow to clearly distinguish and detect point sources from truly diffuse emission even in very dense regions of the sky. With such an instrument we can finally assess complicated regions such as the Galactic center with its supermassive black hole. 
Observations with a GECCO telescope will also shed light on the origin of the Fermi Bubbles, on the origin of the 511 keV line, on the nucleosynthesis of elements and the chemical evolution of the Galaxy, on the dynamics of Galactic winds, on the mechanisms of transport in the low-energy CRs, and eventually on the role of low-energy CRs on the Galaxy evolution and star formation.
Moreover, the possibility of resolving sources at gamma-ray energies will also enable us to answer open questions regarding Galactic diffuse emissions and cosmic rays at large scales. 
In more detail, observations of the diffuse inverse Compton component of the interstellar emission will allow determination of the spatial distribution of low-energy CR electrons, their sources, their propagation and acceleration, and their relation to the interstellar medium. 
As a consequence, a GECCO mission will also enable indirect detection searches for dark matter and searches for new physics \citep[e.g.][]{P8IG} and extragalactic studies \citep[e.g.][]{EGB}. %In particular, it will help in exploring more exotic scenarios and in constraining also other important components such as the extragalactic background light.
Thanks to the power of a GECCO mission to resolve otherwise confused point sources from the diffuse emission and to its unprecedented sensitivity a GECCO mission will also enable studies of single extragalactic and Galactic sources and of populations of sources allowing discoveries of new astrophysical phenomena whose spectra peak in a poorly explored gamma-ray range. With the BGO detector a GECCO mission will also detect transients such as GRBs and will enable improved multimessenger astrophysics.

\acknowledgments
A.M. was supported by NASA award 80GSFC17M0002 and 80NSSC20K0573. E.O. acknowledges the ASI-INAF agreement n. 2017-14-H.0 and the NASA Grant No. 80NSSC20K1558. E.B. acknowledges NASA Grant No. 80NSSC21K0653.  %This is the most common positions for acknowledgments. A macro is
%available to maintain the same layout and spelling of the heading.

%\paragraph{Note added.} This is also a good position for notes added
%after the paper has been written.

%\newline
%\newline
\section*{Author contributions}
E.O. coordinated the paper, provided the contributions on the interstellar emission and cosmic rays, the continuum emission and the de-excitation lines, and on the nucleosynthesis lines; E.B co-coordinated the paper, provided the contributions on the coded-mask mode, on the extragalactic sources, and on the multimessenger synergies; A.M. provided the hardware analysis with the dedicated section on the GECCO mission; A.B. provided the contribution on the Galactic point sources; W.C. provided the contribution on the Compton mode; T.E provided the section on the methodology of separating point sources from diffuse emission; I.M. provided the section on the Galactic center excess; M.N. provided the contribution on the Fermi Bubbles; S.P. provided the contribution related to dark matter and the 511 keV line.


\newcommand{\mnras} {MNRAS}
\newcommand{\apjl}{ApJL}
\newcommand{\apj}{ApJ}
\newcommand{\ssr}{Space Sci. Rev.}
\newcommand{\apjs}{ApJS}
\newcommand{\prd}{PhRvD}
\newcommand{\aap}{A\&A}
\newcommand{\nat}{Nature}
\newcommand{\jcap}{JCAP}
\newcommand{\aaps}{A\&AS}
\newcommand{\prl}{PRL}
\newcommand{\baas}{Bulletin of the American Astronomical Society}
\newcommand{\jgr}{Journal of Geophysical Research}
\newcommand{\apss}{Ap\&SS}
\newcommand{\araa}{Annual Review of Astronomy and Astrophysics}
\newcommand{\nar}{New Astronomy Reviews}

\end{document}